\documentclass[12pt]{article}
\usepackage[pdftex]{graphicx}
\usepackage{amssymb}
\usepackage{amsmath}
\usepackage{bm}
\usepackage{enumerate}

\begin{document}

\begin{titlepage}
	\vskip 2cm
	\begin{center}
		\Large{{\bf Emergent geometry through\\ quantum entanglement in Matrix theories}}
	\end{center}

	\vskip 2cm
	\begin{center}
		{Cameron Gray\footnote{\tt{cbgray@caltech.edu}}, Vatche Sahakian\footnote{\tt{sahakian@hmc.edu}}, William Warfield\footnote{\tt{wwarfield@g.hmc.edu}}}\\
	\end{center}
	\vskip 12pt
	\centerline{\sl Harvey Mudd College}
	\centerline{\sl Physics Department, 241 Platt Blvd.}
	\centerline{\sl Claremont CA 91711 USA}

	\vskip 12pt
	\centerline{\sl California Institute of Technology}
	\centerline{\sl Division of PMA, 1200 E California Blvd}
	\centerline{\sl Pasadena CA 91125 USA}

	\vskip 1cm
	\begin{abstract}

In the setting of the Berenstein-Maldacena-Nastase Matrix theory, dual to light-cone M-theory in a PP-wave background, we compute the Von Neumann entanglement entropy between a probe giant graviton and a source. We demonstrate that this entanglement entropy is directly and generally related to the local tidal acceleration experienced by the probe. This establishes a new map between local spacetime geometry and quantum entanglement, suggesting a mechanism through which geometry emerges from Matrix quantum mechanics. We extend this setting to light-cone M-theory in flat space, or the Banks-Fischler-Shenker-Susskind Matrix model, and we conjecture a new general relation between a certain measure of entanglement in Matrix theories and local spacetime geometry. The relation involves a `c-tensor' that measures the evolution of local transverse area and relates to the local energy-momentum tensor measured by a probe.
	\end{abstract}
\end{titlepage}

\newpage \setcounter{page}{1}

\section{Introduction and highlights}
\label{sec:intro}

Emergent geometry is the general proposition that spacetime is an approximate construct that emerges from underlying collective quantum degrees of freedom. The idea can be made precise within the context string theory~\cite{Ryu:2006ef}-\cite{Nishioka:2009un}, but also arises at low energies and in potentially unrelated settings~\cite{Jacobson:1995ab,Verlinde:2010hp,Li:2010dr}. 

Generically, interactions in a quantum mechanical system lead to entanglement between its interacting parts~\cite{Casini:2009sr}-\cite{peschelcalculation}, a feature that is inherently non-local in character. Entanglement entropy is one measure of this phenomenon but it depends on various choices such as how the physical system is sliced. One realization of the emergent geometry phenomenon posits that, in certain regimes, quantum entanglement can acquire a geometrical flavor -- weaving together what we perceive as the fabric of spacetime. This implies that spacetime geometry is an approximate contraption with quantum entanglement underlying it. 

Gravitational holography seems to play a key role in connecting quantum entanglement and spacetime geometry in the context of string theory. The AdS/CFT, and more generally the bulk/boundary paradigms, use holography to provide a precise dictionary between entanglement and geometry~\cite{Ryu:2006ef}. In this work, we will instead focus on another setting where the holographic mechanism is not obvious, yet a non-gravitational theory describes dual gravitational physics: Matrix theories which are dual to light-cone eleven dimensional supergravity. These include the original Banks-Fischler-Shenker-Susskind (BFSS) Matrix model~\cite{Banks:1996vh} and the newer Berenstein-Maldacena-Nastase (BMN) theory~\cite{Berenstein:2002jq} -- the latter being effectively Matrix theory with an IR regulator, a `Matrix theory in a box'. We will explore the idea of emergent geometry through quantum entanglement in these Matrix theories.

Consider a two-body system, a probe and a source, in BMN theory. In the supergravity dual picture, they interact gravitationally in a PP-wave background. The two objects can be realized as giant gravitons or metastable stars~\cite{Berenstein:2002jq,Dasgupta:2002hx} -- controllable classical solutions of BMN theory. These are spherical membranes moving in the PP-wave background, with a large amount of light-cone momentum for the source membrane, and much smaller for the probe. The source back-reacts on the PP-wave geometry, but, if the two objects are far apart, the local geometry at the probe's location can be a small perturbation of the PP-wave background. On the BMN Matrix theory side, the interaction potential between source and probe can be computed remarkably at weak Yang-Mills coupling at one-loop~\cite{Michelson:2005pk,Lee:2004kv,Wu:2004ma,Lee:2003kf}, yet the result agrees with the supergravity expectations because of an underlying non-renormalization theorem. This BMN computation however makes sense only if the separation between source and probe is large: this regime renders off-diagonal matrix modes heavy, resulting in the needed effective potential for the diagonal modes. This is the setting we will explore to connect quantum entanglement to spacetime geometry. The underlying quantum mechanical system is given by BMN theory, but we start from the outset on the supergravity side with the source and probe being widely separated.

We show that in a two-body interacting system in supergravity, there is a notion of quantum entanglement between the two bodies that is a natural local probe of geometry. Quantum fluctuations about classical trajectories get entangled and the corresponding entanglement entropy is non-trivially geometrical. The connection arises through local tidal forces that the probe feels due to the presence of the source. Furthermore, this link has an intriguing holographic character that refers to a space transverse to both the light-cone direction and the velocity of the probe. We write an explicit relation that connects this entanglement entropy, that can be computed directly in BMN theory, to the local curvature scale near the probe in supergravity. We then generalize this relation -- using symmetry and guesswork -- to a conjecture for a general relation between a certain entanglement entropy of a body with its environment, and the spacetime geometry that the body samples locally\footnote{For a different approach, see~\cite{Anous:2019rqb}}. 

The outline of the paper is as follows. In Section 2.1, the mechanics of tidal acceleration for a probe in light-cone supergravity is presented and applied to the dynamics of the center of mass of a spherical membrane in a PP-wave background in the presence of another spherical membrane. In Section 2.2, a sketch of the one-loop BMN computation is presented, showing the general structure of the interaction potential and the role of membrane dipole interactions. In Section 2.3, we define and compute the Von Neumann entanglement entropy of the gravitating probe, and connect it to the local geometry sampled by the probe. In the last section, the key insights and assumptions are summarized, and a conjecture is made for a general relation between entanglement entropy and spacetime geometry.

\section{Entanglement in a two-body system} 
\label{sec:entang}

\subsection{Supergravity perspective}

Consider a point particle of mass $m$ -- to function as a probe to eleven-dimensional supergravity. It would be described by the standard action
\begin{equation}
	S = \frac{1}{2}\int d\lambda \left(\eta^{-1}g_{\mu\nu}\frac{dx^\mu}{d\lambda}\frac{dx^\nu}{d\lambda} -\eta\, m^2\right)\ ,
\end{equation}
where $\eta$ is the ``vielbein'' on the wordline and $\mu,\nu=0,1,\ldots,10$. We will work in the light-cone frame and hence define
\begin{equation}
	x^\pm = \frac{1}{\sqrt{2}}\left(x^0\pm x^{10}\right)\ .
\end{equation}
The particle is to carry a fixed amount of light-cone momentum.
Furthermore, the background metric $g_{\mu\nu}$ shall be the PP-wave metric $G_{\mu\nu}$ plus a perturbation $h_{\mu\nu}$ that arises from a source present in this PP-wave background; we then write
\begin{equation}
	g_{\mu\nu} = G_{\mu\nu} + h_{\mu\nu}
\end{equation}
The PP-wave background is given by~\cite{Berenstein:2002jq,Wu:2004ma}
\begin{equation}
	G_{++} = -M_{rs} x^r x^s= -M_{ij} x^i x^j-M_{ab} x^a x^b\ \ \ ,\ \ \ G_{+-}=1\ \ \ ,\ \ \ G_{rs}=\delta_{rs}
\end{equation}
with all other components of $G_{\mu\nu}$ being zero. We use the convention 
	\begin{equation}
		i,j, \ldots=1,2,3  \ \ \mbox{ and }\ \  a,b,\ldots=4,\ldots, 9
	\end{equation}
while we reserve $r,s, u, \ldots = 1, \ldots , 9$ for all nine spatial directions. We also define
	\begin{equation}\label{eq:Mij}
		M_{ij} = \frac{\mu^2}{9} \delta_{ij}\ \ \ \mbox{for $i,j=1,2,3$}
	\end{equation}
	\begin{equation}\label{eq:Mab}
		M_{ab} = \frac{\mu^2}{36} \delta_{ab}\ \ \ \mbox{for $a,b=4,\ldots, 9$}
	\end{equation}
where $\mu$ is related to the 4-form flux associated with  the PP-wave background. Hence, the PP-wave breaks the $SO(9)$ symmetry to $SO(3)\times SO(6)$. We proceed with fixing the light-cone gauge by choosing the affine parameter $\lambda=x^+$, and henceforth we use a dot to denote derivative with respect to the light-cone time $x^+$, {\em i.e.} $\dot{x}^r\equiv dx^r/dx^+$. The action becomes
\begin{equation}
	S = \frac{1}{2}\int dx^+ \left(2\,\eta^{-1}\dot{x}^- 
	+\eta^{-1} G_{++} + \eta^{-1} (\dot{x}^r)^2-\eta\, m^2\right) 
	+ \frac{1}{2}\int dx^+\, \eta^{-1} h_{\mu\nu}{\dot{x}^\mu}{\dot{x}^\nu}\ .
\end{equation}
We then have the momentum
\begin{equation}
	p_- = g_{-\mu} p^\mu = p^+ + h_{-\mu}p^\mu  = \frac{\partial L}{\partial \dot{x}^-} = \eta^{-1} + \eta^{-1} h_{-\mu} \dot{x}^\mu
\end{equation}
which implies that $\eta^{-1}$ differs from the light-cone momentum $p^+$ by terms of order $h_{\mu\nu}$. 
The gauge fixing involves the constraint
\begin{equation}
	\frac{\delta S}{\delta \eta} = 0\Rightarrow g_{\mu\nu} \dot{x}^\mu \dot{x}^\nu = -\eta^{2}m^2= -\frac{m^2}{(p^+)^2}+\mathcal{O}(h) \Rightarrow \dot{\tau} = \frac{m}{p^+} +\mathcal{O}(h)
\end{equation}
which determines the relation between the proper time $\tau$ of the probe and light-come time. This constraint can also be used to solve for $\dot{x}^-$ which then can be eliminated from the action in favor of  $h_{\mu\nu}$ and $\dot{x}^r$.

The light-cone Lagrangian is defined by the Routhian
\begin{equation}
	L\rightarrow p_- \dot{x}^- - L
\end{equation}
so that the action takes the form
\begin{equation}\label{eq:sugraaction}
	S = \frac{1}{2}\int dx^+ \left(p^+ G_{++} + p^+ (\dot{x}^r)^2-\frac{m^2}{p^+}\right) - \int dx^+ {V(x,\dot{x})}
\end{equation}
where $V$ collects all $h_{\mu\nu}$-dependent terms that are due to the source which back-reacts on the PP-wave background -- with the $\dot{x}^-$ dependence in $V$ eliminated through the constraint\footnote{$V$ is also averaged over the $x^-$ direction as we will only consider interactions that do not involve light-cone momentum exchange.}. Note also that translation in light-cone time $x^+$ is generated by the light-cone Hamiltonian $p_+$. 

The equations of motion that follow from~(\ref{eq:sugraaction}) take the form 
\begin{eqnarray}
	&&\left(\delta_{rs}+\frac{1}{p^+}\delta_r\delta_s V\right)\ddot{x}^s  = \frac{1}{p^+}\left(-\partial_r V +\partial_s\delta_r V\, \dot{x}^s\right)-M_{rs}x^s  \nonumber \\
	&& \Rightarrow \ddot{x}^r \simeq \frac{1}{p^+}\left(-\partial_r V +\partial_s\delta_r V\, \dot{x}^s\right)-M_{us}x^s \left( \delta_{ru}-\frac{1}{p^+}\delta_r\delta_u V\right)\label{eq:sugraeom1}
\end{eqnarray}
where we define the shorthands
\begin{equation}
	\delta_r V \equiv \frac{\partial V}{\partial \dot{x}^r}\ \ \ ,\ \ \ \partial_r V \equiv \frac{\partial V}{\partial {x}^r}\ .
\end{equation}
We will consider the dynamics to linear order in the metric perturbation, $h_{\mu\nu}$; this means that we can drop powers of the potential $V$ greater than one, which we have done on the second line of~(\ref{eq:sugraeom1}): the back-reaction of the source on the PP-wave geometry is a perturbation, since we will locate the source at the origin of the coordinates and place the probe at a distance far away from it so that the local metric near the probe is close to the PP-wave background's. We can also drop the $\delta_r \delta_u V$ in the last term, $\delta\delta V \ll (p^+)^2$. The  equations of motion become
\begin{equation}\label{eq:sugraeom2}
	\ddot{x}^r \simeq \frac{1}{p^+}\left(-\partial_r V +\partial_s\delta_r V\, \dot{x}^s\right) - M_{rs}x^s\ .
\end{equation}
This last step is more subtle: we will see later on that the entanglement entropy we are interested goes as $\sim V/M_{rs}$; this implies that any $V$-dependent factor that dresses $M_{rs}$ would be sub-leading. 
Note that this is acceleration in light-cone time. For $m\neq 0$, we can relate this to covariant acceleration $a^\mu$ by
\begin{equation}
	\ddot{x}^r \simeq a^r \frac{m^2}{(p^+)^2}
\end{equation}
to linear order in $h$, where $a^r = d^2 x^r/d\tau^2 = du^r/d\tau$. We will see shortly that this level of approximation is suitable. For a massless probe, $m=0$, the affine parameter can be chosen to be $x^+$, so that one can just write $\ddot{x}^r \rightarrow a^r$. 

We next want to introduce a measure of the tidal forces experienced by the probe. To do so, we first define a
space-like vector $\xi^\mu$ that is transverse to velocity. We choose $\xi^+=0$ and write
\begin{equation}
	g_{\mu\nu} \xi^\mu u^\nu = 0 \Rightarrow g_{\mu\nu} \xi^\mu \dot{x}^\nu = 0
\end{equation}
for both massive and massless probes. This then gives
\begin{equation}
	\xi^- \left( 1 + h_{--} \dot{x}^- + h_{-+} + h_{-r} \dot{x}^r\right) +\xi^r \dot{x}^r +h_{rs}\xi^r \dot{x}^s= 0\ .
\end{equation}
We can then solve for $\xi^-$ as needed. Because we have
\begin{equation}
	g_{\mu\nu} \xi^\mu \xi^\nu  = h_{--}\xi^-\xi^-+\xi^r \xi^r + h_{rs} \xi^r \xi^s > 0\ ,
\end{equation}
$\xi^\mu$ is a space-like vector and $\xi^r$ defines a  $11-2=9$ dimensional transverse space to the probe's velocity. 

We denote tidal acceleration by $A^r$ to distinguish it from the probe's covariant acceleration $a^\mu$. Tidal acceleration is generally defined by~\cite{waldbook,Ellis:1984bqf}
\begin{equation}
	A^r = \xi^\nu \nabla_\nu \left(\frac{d^2 x^r}{d\tau^2}\right)\ .
\end{equation}
From this, we are inspired to define tidal light-cone frame acceleration in the PP-wave background as 
\begin{equation}
	A_{PP}^r \equiv \xi^\nu \nabla_\nu \left(\ddot{x}^r + M_{rs} x^s\right)\ .
\end{equation}
We have (a) shifted the acceleration by the effect of the PP-wave background so that this quantity measures exclusively the tidal acceleration from the back-reaction of the source on the PP-wave background; and (b) $A_{PP}$ is acceleration with respect to light-cone time, not proper time, as the dot represents derivative with respect to $x^+$. This means that the expression in the parenthesis is already of order $h_{\mu\nu}$ or equivalently $V$. We can then write
\begin{eqnarray}
	A^r_{PP} && \equiv \xi^\nu \nabla_\nu \left(\ddot{x}^r + M_{rs} x^s\right) = \frac{1}{p^+}\xi^\nu \nabla_\nu \left(-\partial_r V +\partial_s\delta_r V\, \dot{x}^s\right) \nonumber \\
	&& \simeq \frac{1}{p^+}\xi^\nu \partial_\nu \left(-\partial_r V +\partial_s\delta_r V\, \dot{x}^s\right) = \frac{1}{p^+}\xi^u \left(-\partial_u\partial_r V +\partial_u\partial_s\delta_r V\, \dot{x}^s\right) \nonumber \\
	&& \equiv -\frac{1}{p^+}\xi^u K_{ur}
\end{eqnarray}
where we have once again focused on linear order terms in $V$. From this, we can see that, for $m\neq 0$, $A^r_{PP}\simeq A^r m^2/(p^+)^2$ to leading order in $V$, where $A^r$ is defined once again as the tidal acceleration due to the source only, but measured with respect to proper time. For $m=0$, we simply have $A^r_{PP}\equiv A^r$ by choice.

We have now established a relation between the supergravity potential $V$ due to the source and the tidal acceleration $A^r$ experienced by the probe only due to the source with respect to proper time
\begin{equation}\label{eq:ALC1}
	A^r = -\frac{p^+}{m^2}\xi^s K_{rs} \rightarrow -\xi^s K_{rs}\ \ \ \mbox{For $m\neq 0$}\ ,
\end{equation}
\begin{equation}\label{eq:ALC2}
	A^r = -\frac{1}{p^+}\xi^s K_{rs} \rightarrow -\xi^s K_{rs}\ \ \ \mbox{For $m= 0$}\ .
\end{equation}
to linear order in the perturbation due to the source, with
\begin{equation}\label{eq:K1}
	K_{rs} = \partial_r\partial_s V - \partial_s\partial_u\delta_r V\, \dot{x}^u\ ;
\end{equation}
and where we have conveniently normalized $\xi^s$ to get rid of the multiplicative factor.
We will later show that this tensor $K_{rs}$ also determines the quantum entanglement between source and probe. 

The probes we want to focus on shall be the spherical membranes of BMN theory, and we want to track the center of mass motion of such spheres. In supergravity language, the membrane action with worldvolume coordinates $x^+,\theta, \varphi$ is given by~\cite{Lee:2004kv}
\begin{eqnarray}
	L &=& \int d\theta\,d\varphi\, \left[\frac{p^+}{8\pi} \sin\theta \dot{X}^r\dot{X}^r
	-\frac{p^+}{8\pi}\sin\theta M_{rs}X^r X^s
	-\frac{1}{4\pi\,p^+\sin\theta} \left(\partial_\theta X^r\partial_\varphi X^s-\partial_\theta X^s\partial_\varphi X^r\right)^2 \right. \nonumber \\
	&+&\left. \frac{1}{2\pi} \frac{\mu}{3}\epsilon_{ijk} \left(\partial_\theta X^i\right)\left(\partial_\varphi X^j\right)X^k - V(X,\partial X)\right]
\end{eqnarray}
evaluated in the pure PP-wave background. We write
\begin{equation}
	X^i = x^i + Y^i\ \ \ ,\ \ \ X^a = x^a
\end{equation}
with
\begin{equation}
	Y^1 = \mathcal{R}\,\sin\theta \cos\varphi\ \ \ ,\ \ \ 
	Y^2 = \mathcal{R}\,\sin\theta \cos\varphi\ \ \ ,\ \ \ 
	Y^3 = \mathcal{R}\, \cos\theta
\end{equation}
which describes a spherical configuration of radius $\mathcal{R}$. Two special radii result in stable and metastable spherical membranes:
\begin{equation}
	\mathcal{R} = \frac{\mu\,p^+}{6}\ \ \ , \ \ \ m=0\ \ \ \mbox{(Giant graviton)}
\end{equation}
\begin{equation}
	\mathcal{R} = \frac{\mu\,p^+}{12}\ \ \ , \ \ \ m=\frac{(p^+)^2 \mu^2}{72}\ \ \ \mbox{(Metastable star)}
\end{equation}
We only consider these cases. Cross terms between $x$ and $Y$ then integrate to zero so that the center of mass dynamics decouples. We are left with the action  
\begin{equation}
	L = \frac{p^+}{2} \left(\dot{x}^r\right)^2-\frac{p^+}{2}\mu^2 \left(\frac{\left(x^i\right)^2}{9}+\frac{\left(x^a\right)^2}{36}\right) -\frac{m^2}{p^+} - {V(x,\dot{x})}\label{eq:potsugra}
\end{equation}
where in the last step we added the perturbation of the PP-wave background by a source through the addition of a potential $V$. We see that our prior analysis is validated with two possible values of the mass $m$, zero or ${(p^+)^2 \mu^2}/{72}$. However, $V$ now includes interactions due to membrane charge {\em in addition to} gravitational forces. In the regime of interest where the spherical membranes are much smaller than the distance between them, one can still have forces arising from membrane dipole-dipole interaction. Each spherical membrane would have zero net membrane charge, yet would carry a dipole that leads to a force due to the non-uniform four-form flux from the other. Note however that the flux from the background PP-wave does not impact the center of mass dynamics of the membranes because this flux is uniform. Hence, $V$ includes gravitational and membrane-dipole effects, while PP-wave background effects are absent from $V$ and arise only in the middle $\mu$-dependent terms of~(\ref{eq:potsugra}).

In the spirit of treating our action as that of a probe, the source (henceforth referred to as object 2) is to be `heavy' compared to the `probe'  (referred to as object 1). In light-cone frame language, this implies the light-cone momentum of the source, $p^+_2$, should be much larger than light-cone momentum of the source, $p^+_1$
\begin{equation}
	p^+_2\gg p^+_1\ .\label{eq:probelimit}
\end{equation}
	We place the source at the origin $x_2^r=0$ of the PP-wave background coordinates, and we can see from the equations of motion~(\ref{eq:sugraeom2}) that the motion of the source would be negligible in relation to that of the probe. We place the probe a distance far away $x^r=x^r_1-x^r_2$. We can then use all the relations for the point probe we presented and apply them to the membrane probe's center of mass motion, with $p^+_1\rightarrow p^+$ and $x_1^r \rightarrow x^r$. The light-cone momentum of the source, $p_2^+$, would then only appear in the potential $V$.

\subsection{Computing $V$ in BMN theory}
\label{sec:giant}

In this section, we sketch the derivation of the effective potential $V$ discussed in the previous section, but from the perspective of the dual BMN theory. The computation is already done in the literature at one-loop level~\cite{Kabat:1997im,Becker:1997xw,Lee:2003kf,Wu:2004ma,Lee:2004kv} so we will only present an abbreviated version. However, we will include new terms that account for couplings to membrane dipole charge that have not been presented. These new terms are interesting for us as they extend the map between geometry and entanglement entropy, as we shall see. While the BMN computation is done at weak coupling, the results are expected to carry over to strong coupling and hence match with the supergravity potential due to a renormalization theorem. Indeed, it has been shown that the one-loop potential from BMN theory agrees with the supergravity expectations in several settings\footnote{Note however than terms that depends on membrane charge have not been compared.}.

BMN theory is $0+1$ dimensional $U(N)$ Super Yang-Mills (SYM) theory that is purported to be dual to
light-cone gauge M-theory in a PP-wave background. The rank of the gauge group $N$ maps onto light-cone momentum in M-theory. 
Our starting point is the Matrix theory action in the background field gauge\footnote{We will try to follow, as much as possible, the notation and conventions used in~\cite{Kabat:1997im} and~\cite{Sugiyama:2002bw}.}
\begin{eqnarray}
	S&=& \int dt\, \mbox{Tr} \left[
	\frac{1}{2R}D_t X^r\, D_t X^r +\frac{R}{4} [X^r,X^s]^2 +  \Psi_\alpha D_t \Psi_\alpha +i\, R\, \Psi_\alpha \Gamma^r_{\alpha\beta} [X^r,\Psi_\beta] \right. \nonumber \\
	&-& \left.  \frac{1}{2R}\frac{\mu^2}{9} X^i X^i -\frac{1}{2R}\frac{\mu^2}{36} X^a X^a - \frac{i\,\mu}{3} \epsilon_{ijk} X^iX^jX^k-\frac{\mu}{4}\Psi_\alpha\Gamma^{123}_{\alpha\beta}\Psi_\beta
	 \right. \nonumber \\
	&-& \left. \frac{1}{2}(\partial_t A+i[X^r_{\mathrm{bg}},X^r])^2  + \frac{i}{2} \partial_t\overline{G} \left(\partial_t G-i \left[A,G\right]\right)-\frac{1}{2}\overline{G}[X^r_{\mathrm{bg}},[X^r,G]]
	\right]\ .\label{eq:matrixtheory}
\end{eqnarray}
$D_t \equiv \partial_t -i\,\left[A, \cdot \right]$. We use units such that the eleven dimensional Planck length is set to one $\ell_P=1$.
All fields are in the adjoint of $U(N)$, and the spinor fields $\Psi_\alpha$ are $10$ dimensional Majorana-Weyl.
The background field gauge condition requires~\cite{Abbott:1981ke}
\begin{equation}
	\partial_t A+i[X^r_{\mathrm{bg}},X^r] = 0\ ,
\end{equation}
and $G$ is a matrix of Faddeev-Popov ghosts. $R$ is the radius of the M-theory light-cone circle. It functions as the Yang-Mills coupling of the theory. 
We take the background as
\begin{equation}
	X^r_{\mathrm{bg}} = \left(
	\begin{array}{cc}
		{X}^r_1(t) & 0 \\
		0 & {X}^r_2(t)
	\end{array}
	\right)
\end{equation}
with all other fields vanishing.
This is a block diagonal configuration with ${X}^i_1$ being an $N_1\times N_1$ matrix, and ${X}^i_2$ being an $N_2\times N_2$ matrix; we have $N=N_1+N_2$. In M-theory language, ${X}^i_1$ is to represent an object that carries $N_1$ units of light-cone momentum -- such as a giant graviton with $p^+_1=N_1/R$; while ${X}^i_2$ represents another object with $N_2$ units of light-cone momentum. At the end, we will choose ${X}^i_1$ and ${X}^i_2$ to describe spherical membranes, giant gravitons or metastable stars.

To perform the path integrals, we use the Euclidian form with $t\rightarrow i\, \tau$ and $A\rightarrow -i\, A$. We also rescale our parameters as follows: $t\rightarrow t/R$, $A\rightarrow A\,R$, $\mu\rightarrow \mu\,R$.

We then want to write down an effective action by perturbing this background by
\begin{equation}
	\begin{array}{lll}
		A_0 = \left(
		\begin{array}{cc}
			a_1(t) & a(t) \\
			\overline{a}(t) & a_2(t)
		\end{array}
		\right) &
		X^r = X^r_{\mathrm{bg}} + \left(
		\begin{array}{cc}
			x_1^r(t) & x^r(t) \\
			{x}^{r\,\dagger}(t) & x_2^r(t)
		\end{array}
		\right) & 
		\Psi_\alpha = \left(
		\begin{array}{cc}
			\psi_{1\alpha}(t) & \psi_\alpha(t) \\
			{\psi^\dagger}_\alpha(t) & \psi_{2\alpha}(t)
		\end{array}
		\right)
	\end{array}\label{eq:perturbations}\ .
\end{equation}
We henceforth refer to $a_1$, $a_2$, $x_1^r$, $x_2^r$, $\psi_{1\alpha}$ and $\psi_{2\alpha}$ as {\em diagonal} fluctuations or modes; while the other fluctuations are said to be {\em off-diagonal}.
The centers of mass of the two background objects are given by
\begin{equation}
	\overline{x}_{1,2}^r \equiv \frac{\mbox{Tr}\, {X}_{1,2}^r}{N_{1,2}}\ ,
\end{equation}
where we use the bar notation to denote variables which are not matrices;
while the size of each object might naturally be represented by the second moments
\begin{equation}
	\mathcal{R}_{1,2}^2 \equiv \frac{\mbox{Tr}\, ({X}_{1,2}^r)^2}{N_{1,2}} - (\overline{x}_{1,2}^r)^2\ .
\end{equation}
We assume that the two background objects are widely separated from each other. In this regime, the off-diagonal perturbations in~(\ref{eq:perturbations}) are heavy or high frequency modes. One can then integrate them out and derive the effective potential for the background variables ${X}_1^r$ and ${X}_2^r$.

To be more precise, we substitute~(\ref{eq:perturbations}) into~(\ref{eq:matrixtheory}) and expand in the small perturbations about the background. We then note the following:
\begin{itemize}
	\item At quadratic order in the perturbations, there are no couplings between off-diagonal and diagonal perturbations; the two sectors decouple. 
	\item The full action is quadratic in the fermions and, along with the previous observation, this implies that there are no couplings between off-diagonal and diagonal fermionic perturbations at {\em any} order. Hence, fermions on the diagonal do not talk to off-diagonal fermions, except at higher orders through interactions with bosonic perturbations. 
	\item The diagonal perturbations do not involve any couplings between the two objects, {\em i.e.} no $x_1 x_2$ terms arise. This implies that the leading contribution to the effective potential between the two objects does not get a contribution from fluctuations on the diagonals.
	\item The frequency of off-diagonal modes scale with $\mbox{Tr} (K^r)^2$ where
	\begin{equation}\label{eq:Kr}
		K^r \equiv {X}_{1}^r \otimes \bm{1}_{N_2 \times N_2} -  \bm{1}_{N_1 \times N_1}\otimes ({X}_{2}^{r})^T
	\end{equation}
	which can be deemed as a matrix distance between the two background objects.
\end{itemize}

These observations allow us to set the diagonal perturbations to zero, and integrate out the off-diagonal modes -- expanded to quadratic order in a regime where they are heavy, {\em i.e.} when the timescale of evolution of the background matrices is much longer than the timescale of oscillations from the off-diagonal perturbations. This yields the one-loop effective potential for the background matrix configuration. The off-diagonal perturbations consist of $10\, N_1 N_2$ bosonic modes with frequency squared matrix $\mathcal{M}_b$, $16\, N_1 N_2$ fermionic modes with frequency squared matrix $\mathcal{M}_f$, and $2\,N_1 N_2$ ghosts with $\mathcal{M}_g$. The effective action then becomes
\begin{equation}\label{eq:Sv}
	V = -\int dt\, \left(\mbox{Tr} \sqrt{\mathcal{M}_{b}}-\frac{1}{2} \mbox{Tr} \sqrt{\mathcal{M}_{f}}
	-2\,\mbox{Tr} \sqrt{\mathcal{M}_g}\right)\ .
\end{equation}
We write $\mathcal{M}_b=\mathcal{M}_{0b}+\mathcal{M}_{1b}$ and $\mathcal{M}_f=\mathcal{M}_{0f}+\mathcal{M}_{1f}$ with
\begin{align*}
	\mathcal{M}_{0b} &= \sum_r K^{r\,2}\otimes \bm{1}_{10\times 10} \\
	\mathcal{M}_{1b} &= \left(
\begin{array}{ccc}
 0 & 2 i \partial_t K^j & 2 i  \partial_t K^b \\
 -2 i \partial_t K^i & 2 [K^i,K^j]+\frac{\mu ^2 }{9}\delta^{ij}-i \mu  \epsilon^{ijk} K^k & 2 [K^i,K^b] \\
 -2 i \partial_t K^a & 2 [K^a,K^j] & 2 [K^a,K^b]+\frac{\mu ^2 }{36} \delta^{ab}\\
\end{array}
\right)
\end{align*}
\begin{align*}
	\mathcal{M}_{0f} &= \sum_r K^{r\,2}\otimes \bm{1}_{16\times 16} \\
	\mathcal{M}_{1f} &= \partial_t K^r \otimes \Gamma^r + \frac{1}{2}\left[K^r,K^s\right]\otimes\Gamma ^{rs}-\frac{1}{4} i \mu  \epsilon ^{ijk} K^i\otimes\Gamma^{jk}+\frac{\mu ^2}{16}
\end{align*}
and
\begin{equation}
	\mathcal{M}_g = \sum_r K^{r\,2}\ .
\end{equation}
where $K^r$ was defined above in~(\ref{eq:Kr}).
We then use the integral representation for the square root of a matrix using a Dyson series, {\em e.g.}
\begin{eqnarray}
	 && \left. \mbox{Tr} \sqrt{\mathcal{M}_{0b}+\mathcal{M}_{1b}} = -\frac{1}{2\sqrt{\pi}}  \mbox{Tr} \int_0^\infty \frac{d\tau}{\tau^{3/2}} e^{-\tau(\mathcal{M}_{0b}+\mathcal{M}_{1b})} \right. \nonumber \\
	 && \left. = -\frac{1}{2\sqrt{\pi}} \mbox{Tr} \Big(  
	 \int_0^\infty \frac{d\tau_1}{\tau_1^{3/2}} e^{-\tau_1 \mathcal{M}_{0b}} \mbox{Tr}_\text{L} (1) -  \int_0^\infty \int_0^\infty \frac{d\tau_1 d\tau_2}{(\tau_1+\tau_2)^{3/2}} e^{-(\tau_1+\tau_2 )\mathcal{M}_{0b}} \mbox{Tr}_\text{L} (\mathcal{M}_{1b}(\tau_2) )  \right. \nonumber \\
	 && \left.  + \int_0^\infty  \int_0^\infty \int_0^\infty \frac{d\tau_1 d\tau_2 d\tau_3}{(\tau_1+\tau_2 \tau_3)^{3/2}} e^{-(\tau_1 +\tau_2+\tau_3)\mathcal{M}_{0b}} \mbox{Tr}_\text{L} (\mathcal{M}_{1b}(\tau_2+\tau_3) \mathcal{M}_{1b}(\tau_3)) + ...   \Big) \right. \nonumber \\
\end{eqnarray}
where we defined
\begin{equation}
	\mathcal{M}_1(\tau) \equiv e^{\tau\, \mathcal{M}_0}\, \mathcal{M}_1\, e^{-\tau\, \mathcal{M}_0}\ .
\end{equation}
$\mbox{Tr}_L$ involves tracing over Lorentz space, while $\mbox{Tr}$ refers to tracing over group space.
We keep terms up to third order in this Dyson series. We evaluate the Dyson integrals in the regime where the distance between the center of masses of the two objects, related to $(K^r)^2$, is large so that the small $\tau$ region of the integrands dominates~\cite{Kabat:1997im}).

At zeroth order, the zero point energies cancel as expected: $10N_1N_2-(1/2)(16 N_1N_2)-2 N_1N_2=0$. At first order, the cancellation continues
\begin{equation}
\mbox{Tr} (\mathcal{M}_{1b}) -\frac{1}{2}\mbox{Tr}(\mathcal{M}_{1f}) =\frac{\mu^2}{2}-\frac{\mu^2}{2}=0\ .
\end{equation}
At second order, one gets the first nonzero term
\begin{equation}
\mbox{Tr} (\mathcal{M}_{1b} \mathcal{M}_{1b}) - \frac{1}{2}\mbox{Tr}(\mathcal{M}_{1f} \mathcal{M}_{1f}) =\frac{\mu^4}{96}
\end{equation}
And at third order the expression grows considerably
\begin{eqnarray}
     \mbox{Tr} (\mathcal{M}_{1b} \mathcal{M}_{1b}\mathcal{M}_{1b}) -\frac{1}{2}\mbox{Tr}(\mathcal{M}_{1f} \mathcal{M}_{1f}\mathcal{M}_{1f}) &=& - \frac{7}{6}\mu^2 F_{0a}^2 - \frac{1}{6}\mu^2 F_{0i}^2 - \frac{5}{12}\mu^2 F_{ab}^2 + \frac{1}{6}\mu^2 F_{ai}^2 + \frac{7}{12}\mu^2 F_{ij}^2 \nonumber \\
     &+& \frac{7}{12}\mu^3 \epsilon_{ijk} K_i F_{jk} + \frac{7}{24}\mu^4 K_i^2+\frac{95}{41472} \mu^6
\end{eqnarray}
where we have defined $F_{0r} = \partial_t K_r$ and $F_{rs} =  i[K_r,K_s]$ to make things concise.

Putting things together, we obtain the general expression
\begin{eqnarray}
    V &=& \mbox{Tr}\left[ - \frac{1}{768\,R^3}\frac{\mu^4}{\overline{x}^3} -\frac{1}{16\,R^3}\frac{\mu^2}{\overline{x}^5}\left(\frac{7}{6} F_{0a}^2 + \frac{1}{6}F_{0i}^2\right)   + \frac{7}{384\,R^3}\frac{\mu^4}{\overline{x}^5} K_i^2 \right.\nonumber \\
   & & \left. \left. - \frac{1}{16\,R}\frac{\mu^2}{\overline{x}^5}\left(\frac{5}{12}F_{ab}^2 - \frac{1}{6}F_{ai}^2 - \frac{7}{12}F_{ij}^2\right)  + \frac{7}{192\,R^2}\frac{\mu^3}{\overline{x}^5} \epsilon_{ijk} K_i F_{jk} + \frac{95}{663552\,R^5}\frac{\mu^6}{\overline{x}^5}   \right. \right] \label{eq:VV1}\nonumber \\
\end{eqnarray}
where we have restored the factors of $R$ to undo the rescaling of time and $\mu$. This expression is then written in units where $\ell_P=1$, as was the original action~(\ref{eq:matrixtheory}). The first line matches with the result of~\cite{Lee:2003kf}. The terms on the second line of~(\ref{eq:VV1}) include commutators of the matrices which are non-zero when the size and shape of each object is taken into account. These terms have not been presented in the literature for BMN theory to our knowledge and involve effects from coupling to membrane charge. To be more specific, we next substitute for spherical membrane configurations 
\begin{align*}
    {X}_1^i &= \overline{x}_1^i + \frac{\mu}{3} J_1^i \\
    {X}_1^a &= \overline{x}_1^a \\
    {X}_2^i &= \overline{x}_2^i + \frac{\mu}{3} J_2^i \\
    {X}_2^a &= \overline{x}_2^a
\end{align*}
where the $SU(2)$ generators are written as $[J^i, J^j] = i \epsilon_{ijk} J^k$ with Casimir ${\mathrm{Tr}(J_i^2)} = N{(N^2 - 1)}/{4}$. We are considering the stable giant gravitons as an example. We then have 
\begin{equation}
    {K}^i = \overline{x}^i \bm{1}_{N_1 \times N_1} \otimes \bm{1}_{N_2 \times N_2} + \frac{\mu}{3} J_1^i \otimes \bm{1}_{N_2 \times N_2} - \frac{\mu}{3} \bm{1}_{N_1 \times N_1} \otimes (J_2^i)^T
\end{equation}
\begin{equation}
	{K}^a = \overline{x}^a \bm{1}_{N_1 \times N_1} \otimes \bm{1}_{N_2 \times N_2}
\end{equation}
where 
\begin{equation}
	\overline{x}^r \equiv \overline{x}_1^r - \overline{x}_2^r
\end{equation}
is the distance between the center of masses. Tracing over the color space, we get an effective potential between giant gravitons given by
\begin{eqnarray}
    V &=& -\frac{\mu^4}{768\,R^3}\frac{N_1 N_2}{\overline{x}^3} -\frac{\mu^2}{96\,R^3} \frac{N_1 N_2}{\overline{x}^5}\left((\dot{\overline{x}}^i)^2 +7\,(\dot{\overline{x}}^a)^2\right) + \frac{7\mu^4}{384\,R^3}\frac{N_1 N_2}{\overline{x}^5} (\overline{x}^i)^2  \nonumber \\
    &+&  \frac{61 \mu^6}{1990656\,R^5}\frac{N_1 N_2}{\overline{x}^5}  + \frac{7\mu^6}{124416\,R^5}\frac{N_1 N_2 (N_1^2+N_2^2)}{\overline{x}^5}\label{eq:finpot}\ .
\end{eqnarray}
Once again, the second line includes contributions  from the commutators of objects and hence incorporate their mutual interaction through dipole membrane charge\footnote{We note that there can in principle be higher order terms that mix with the terms shown at this order. We can see this in the third term already where what naively appears as an $\overline{x}^{-5}$ order term in the large distance expansion can actually contribute at the $\overline{x}^{-3}$ order. This pattern continues in the large $\overline{x}$ expansion.} 

To relate this expression to the potential in~(\ref{eq:potsugra}), we identify the light-cone momenta $p^+_{1,2}=N_{1,2}/R$, and we take the probe limit where $N_1 \ll N_2$, object 1 being the probe and object 2 being the source, along~(\ref{eq:probelimit}). We then have
\begin{eqnarray}
    {V} &=& p_1^+p_2^+ \left[-\frac{\mu^4}{768\,R}\frac{1}{\overline{x}^3} -\frac{\mu^2}{96\,R} \frac{1}{\overline{x}^5}\left((\dot{\overline{x}}^i)^2 +7\,(\dot{\overline{x}}^a)^2\right) + \frac{7\mu^4}{384\,R}\frac{(\overline{x}^i)^2}{\overline{x}^5} \right.  \nonumber \\
    &+& \left. \frac{61 \mu^6}{1990656\,R^3}\frac{1}{\overline{x}^5}  + \frac{7\mu^6\,(p_2^+)^2}{124416\,R}\frac{1}{\overline{x}^5}\right]\label{eq:finpot}\ .
\end{eqnarray}
We will comment later on the significance of the non-zero contribution from dipole membrane interactions appearing on the second line. Restoring the Planck length, the first line is multiplied by $\ell_P^9$, while the second one involves $\ell_P^{15}$, with the eleven dimensional gravitational constant given by $\kappa_{11}^2= 16\,\pi^5\ell_P^9$. With the tension of the membrane scaling as $T_2\sim 1/\ell_P^3$, this corresponds to $\kappa_{11}^4\,T_2$.

\subsection{Entanglement between two interacting objects}

We consider two objects interacting with a translationally and rotationally invariant potential, and we want to compute a certain measure of quantum entanglement entropy between them due to their interaction. The computation is quite general, but we will gradually narrow it down to the case at hand: a probe moving far away from a source in light-cone PP-wave background.

The light-cone action for the two body system is given by
\begin{eqnarray}
	S_{PP} &=& \int dx^+ \left(\frac{p^+_1}{2}\dot{x}_1\cdot\dot{x}_1+\frac{p^+_2}{2}\dot{x}_2\cdot\dot{x}_2\right. \nonumber \\
	&-&\left.\frac{p_1^+}{2}{x}_1\cdot M\cdot {x}_1-\frac{p_2^+}{2}{x}_2\cdot M\cdot {x}_2-V(x_1-x_2,\dot{x}_1-\dot{x}_2) \right)
\end{eqnarray}
with $x\cdot x\equiv x^r x^r$ where $r=1,\ldots, 9$. Once again, time and the dot notation refer to light-cone time $x^+$. $M$ will be related to the matrix encountered in equations~(\ref{eq:Mij}) and~(\ref{eq:Mab}) from the PP-wave background. We next note the following about the system of interest:
\begin{itemize}
	\item $V$ is the effective potential that arises from supergravity or, equivalently, BMN theory. By symmetry, it depends only on the relative coordinates $x^r \equiv x_1^r-x_2^r$ and $\dot{x}^r = \dot{x}_1^r-\dot{x}_2^r$, and is rotationally invariant. Furthermore, $V$ is even in $\dot{x}$ because of time reversal symmetry.
	\item The action is written in the light-cone frame, with Lorentz symmetry broken to Galilean symmetry as expected. 
	\item Given the symmetries, the structural form of $V$ is 
	\begin{equation}
		V \rightarrow V(\dot{x}\cdot \dot{x}, \dot{x}\cdot x, {x}\cdot x)\ .
	\end{equation}
	\item We will assume that the two objects are separated by large enough distances that the potential $V$ is a small perturbation to their dynamics.
	\item $M$ is present to reproduce the effect of the PP-wave background.
\end{itemize}

This two-body system has classical solutions given by $X^r(t)$ for given initial conditions; and we consider perturbing such a solution 
\begin{equation}
	x_{1,2}^r=X_{1,2}^r+\varepsilon_{1,2}^r
\end{equation}
and write the corresponding potential expanded to quadratic order
\begin{eqnarray}
	V &=& V(X,\dot{X}) + \frac{1}{2}\left(
	\varepsilon_1^r-\varepsilon_2^r\right)\left(
	\varepsilon_1^s-\varepsilon_2^s\right)\partial_r\partial_s V  \nonumber \\
	&+&\frac{1}{2}\left(
	\dot{\varepsilon}_1^r-\dot{\varepsilon}_2^r\right)\left(
	\dot{\varepsilon}_1^s-\dot{\varepsilon}_2^s\right)\delta_r\delta_s V
	+\left(
	{\varepsilon}_1^r-{\varepsilon}_2^r\right)\left(
	\dot{\varepsilon}_1^s-\dot{\varepsilon}_2^s\right)\partial_r\delta_s V
\end{eqnarray}
where $X^r\equiv X_1^r-X_2^r$. Note that $X$ here is {\em not} a matrix, as it was in the previous section. Given that we are perturbing around a solution to the equations of motion, the linear terms cancel. All derivatives of $V$ are to be evaluated at the classical time-dependent solution. We write $\varepsilon \equiv \varepsilon_1-\varepsilon_2$ and rearrange the last term as follows
\begin{equation}
	\varepsilon^r\dot{\varepsilon}^s \partial_r\delta_s V = -\frac{1}{2} \varepsilon^r \varepsilon^s \dot{X}^\alpha \partial_r \partial_\alpha \delta_s V-\frac{1}{2} \dot{\varepsilon}^r \varepsilon^s \left(\partial_r \delta_s V-\partial_s \delta_r V\right)\ .
\end{equation}
We then define antisymmetric tensor
\begin{equation}\label{eq:Brs}
B_{rs} \equiv \partial_r \delta_s V-\partial_s \delta_r V	\ .
\end{equation}
Note that, given the expected form of $V$ from above, nonzero contributions to $B$ must take the form
\begin{equation}
	B_{rs} \propto X^r \dot{X}^s - X^s \dot{X}^r
\end{equation}
which consist of the angular momenta of the classical solution. We next define
\begin{equation}\label{eq:Krs}
	K_{rs}\equiv \partial_r\partial_s V - \dot{X}^\alpha \partial_\alpha \partial_r \delta_s V 
\end{equation}
which is the same expression encountered in~(\ref{eq:K1}).
We then write the potential as
\begin{equation}
	V \rightarrow  \frac{1}{2}\left(
	\varepsilon_1^r-\varepsilon_2^r\right)\left(
	\varepsilon_1^s-\varepsilon_2^s\right)K_{rs}	+\frac{1}{2}\left(
	\dot{\varepsilon}_1^r-\dot{\varepsilon}_2^r\right)\left(
	\dot{\varepsilon}_1^s-\dot{\varepsilon}_2^s\right)\delta_r\delta_s V
	-\frac{1}{2}\left(
	{\varepsilon}_1^r-{\varepsilon}_2^r\right)\left(
	\dot{\varepsilon}_1^s-\dot{\varepsilon}_2^s\right)B_{rs}
\end{equation}
dropping constant terms. Putting things back together, the Lagrangian of the perturbed system becomes
\begin{eqnarray}
	L &=& \frac{p_1^+}{2} \dot{\varepsilon}_1\cdot \dot{\varepsilon}_1
+\frac{p_2^+}{2} \dot{\varepsilon}_2\cdot \dot{\varepsilon}_2 - \frac{1}{2}(\dot{\varepsilon}_1^r-\dot{\varepsilon}_2^r) (\dot{\varepsilon}_1^s-\dot{\varepsilon}_2^s)\, \delta_r\delta_s V \nonumber \\
&-& \frac{p_1^+}{2}{\varepsilon}_1\cdot M\cdot {\varepsilon}_1
- \frac{p_2^+}{2}{\varepsilon}_2\cdot M \cdot {\varepsilon}_2 - \frac{1}{2}(\varepsilon_1^r-\varepsilon_2^r)(\varepsilon_1^s-\varepsilon_2^s) K_{rs} \nonumber \\
&+& \frac{1}{2}\left(
	{\varepsilon}_1^r-{\varepsilon}_2^r\right)\left(
	\dot{\varepsilon}_1^s-\dot{\varepsilon}_2^s\right)B_{rs}
\end{eqnarray}
We can diagonalize the kinetic terms perturbatively given that the potential $V$ is small
\begin{equation}
	\varepsilon_1^r = \epsilon_1^r-\epsilon_2^s \frac{1}{p_1^+-p_2^+}\delta_r \delta_s V\ \ \ ,\ \ \ \varepsilon_2^r = \epsilon_2^r+\epsilon_1^s \frac{1}{p_1^+-p_2^+}\delta_r \delta_s V
\end{equation}
yielding
\begin{eqnarray}
	L &=& \frac{1}{2}\left(p_1^+ \delta_{rs}-\delta_r \delta_s V\right)\dot{\epsilon}_1^r\dot{\epsilon}_1^s + \frac{1}{2}\left(p_2^+ \delta_{rs}-\delta_r \delta_s V\right)\dot{\epsilon}_2^r\dot{\epsilon}_2^s \nonumber \\
	&-&\frac{1}{2}M_{rs} \left(p_1^+\epsilon_1^r \epsilon_1^s+p_2^+\epsilon_2^r \epsilon_2^s\right) \nonumber \\
	&+&\frac{1}{2}K_{rs}\left(\epsilon_1^r \epsilon_2^s+\epsilon_1^s \epsilon_2^r\right) - \frac{1}{2}\left(
	{\epsilon}_1^r\dot{\epsilon}_2^s+{\epsilon}_2^r\dot{\epsilon}_1^s\right)B_{rs}\ .
\end{eqnarray}
Several assumptions underly this last expression. First, we have dropped time derivatives of the potential in relation to time derivatives of the perturbations -- assuming an adiabatic regime where the time scale of evolution of the perturbations is much shorter than the time scale associated with the classical trajectories which determine the evolution of the potential. At the end of this section, we will revisit this issue and justify this approximation. We have also assumed that the potential $V$ is small -- treating the back-reaction from the source onto the PP-wave background small. This justifies the diagonalization of the kinetic terms to leading linear order in $V$; and also dropping $K_{rs}$ and $B_{rs}$ in relation to $M_{rs}$ that arises in expressions  that couple object 1 to itself and object 2 to itself. The latter step is somewhat subtle: given that $K_{rs}$ and $B_{rs}$ are evaluated at a classical solution, and considering sources and probes that conform to the $SO(3)\times SO(6)$ symmetry of the PP-wave background -- such as giant gravitons and metastable stars, the relevant perturbations for the computation of entanglement will lie either in $SO(3)$ or $SO(6)$, as we shall soon see; this means that $M_{rs}$ is effectively proportional to the identity. As a result, terms like $p_1^+ M_{rs}+K_{rs}$ can be approximated consistently as just $p_1^+ M_{rs}$: simply diagonalize $K_{rs}$, drop the eigenvalues in relation to the PP-wave background, then retransform back -- leaving $M_{rs}$ unchanged. All this brings us to the expression above. We now note that we can also drop the $\delta_r\delta_s V$ in relation to the $\delta_{rs}$ for the same reasons outlined in Section 2.1. 
Finally, we rescale the perturbation variables conveniently as
\begin{equation}
	z^r_{1,2}\equiv \epsilon^r_{1,2}\sqrt{p_{1,2}^+}
\end{equation}
so that the Lagrangian takes the more canonical form
\begin{eqnarray}
		&&L \simeq \frac{1}{2} \dot{z}_1^2 +\frac{1}{2} \dot{z}_2^2 - \frac{1}{2} M_{rs}\left({z_1^r z_1^s}+{z_2^r z_2^s}\right) + \frac{1}{2}\frac{{K_{rs}}}{\sqrt{p_1^+p_2^+}}\left(z_1^r z_2^s+ z_1^s z_2^r\right) \nonumber \\
		&& -\frac{1}{2}\frac{B_{rs}}{\sqrt{p_1^+p_2^+}}\left(
	{z}_1^r\dot{z}_2^s+{{z}_2^r}\dot{z}_1^s\right)\ .\label{eq:Llong}
\end{eqnarray}
$z_1$ and $z_2$ are the perturbations of the classical solution and they describe, to this order of the treatment, quadratic oscillators. Note however that the frequencies involved depend on the classical trajectory and hence are in general time dependent. However, as alluded to above, in a proper adiabatic regime where the time scale of evolution of the oscillators is much shorter than the time scale associated with the classical trajectories, the quantum state for the $z_1$-$z_2$ system should be taken to be the ground state of the simple harmonic oscillators. This ground state will in general involve entanglement between $z_1$ and $z_2$, and it is the entropy associated with this entanglement that we want to next compute as a measure of entanglement between objects 1 and 2. At the end, we will also come back to analyze the validity of the assumed adiabatic regime.

To go further, we treat two different scenarios separately: one where $B_{rs}=0$, and then a more general case.

\vspace{0.5cm}{\bf Radial motion}\vspace{0.5cm}

We first consider boundary conditions for the classical trajectories such that all angular momenta vanish; that is, we consider radial motion. We then have
\begin{equation}
	B_{rs}=0\ \ \ \mbox{For zero angular momentum}
\end{equation}
Note also that, given the expected $SO(3)\times SO(6)$ symmetry of the system we will eventually map onto, this radial motion will live either in $SO(3)$ or $SO(6)$, and not the more general $SO(9)$. The Lagrangian can then be written in direct product matrix form as
\begin{equation}
		L = \frac{1}{2}\dot{z}\cdot \dot{z}-\frac{1}{2} z\cdot \hat{\kappa}\cdot z 
\end{equation}
where
\begin{equation}
	z^r = (z^r_1, z^r_2)\ ;
\end{equation}
and we define
\begin{equation}
	\hat{\kappa} = 
		M_{rs} \otimes\left(
	\begin{array}{cc}
		1 & 0 \\
		0 & 1
	\end{array}
	\right) + 
		{\frac{K_{rs}}{\sqrt{p_1^+p_2^+}}}\otimes\left(
	\begin{array}{cc}
		0 & -1 \\
		-1 & 0
	\end{array}
	\right)
\end{equation}
The ground state wavefunction for the system is then a gaussian given by~\cite{Casini:2009sr}
\begin{equation}
	\psi(z_1,z_2) = \left(\mbox{det} \frac{\hat{\Omega}}{\pi}\right)^{1/4} \exp\left[{-\frac{1}{2}z^T\cdot \hat{\Omega}\cdot z}\right]
\end{equation}
where
\begin{equation}\label{eq:omega}
	\hat{\Omega} = \sqrt{\kappa} \equiv \left(
	\begin{array}{cc}
		(\Omega_{11})_{rs} & (\Omega_{12})_{rs} \\
		(\Omega_{21})_{rs} & (\Omega_{22})_{rs}
	\end{array}
	\right)\ .
\end{equation}
To compute the quantum entanglement between the two bodies in this quantum state, we write the reduced density matrix~\cite{Casini:2009sr}
\begin{equation}
	\rho_1(z_1,z'_1) = \mbox{Tr}_2 \left[ \rho \right] = \sqrt{\mbox{det}\frac{1-\hat{C}}{\pi}}e^{-\frac{1}{2}Z_1\cdot Z_1} e^{-\frac{1}{2}Z_1'\cdot Z_1'} e^{\frac{1}{4}(Z_1+Z_1')\cdot \hat{C} \cdot (Z_1+Z_1')}
\end{equation}
where
\begin{equation}
	Z_1^r = (\hat{\Omega}_{11}^{1/2})_{rs} z_1^s
\end{equation}
and
\begin{equation}\label{eq:lambda}
	\hat{C} \equiv \hat{\Omega}_{11}^{-1/2}\cdot\hat{\Omega}_{12}\cdot\hat{\Omega}_{22}^{-1}\cdot\hat{\Omega}_{21} \cdot\hat{\Omega}_{11}^{-1/2}
\end{equation}
Note that the latter is a $9$ by $9$ matrix living in the space transverse to the relative motion. The Von Neumann entropy of interest is then given by~\cite{Casini:2009sr}
\begin{eqnarray}
	S_{ent} &=& \mbox{Tr} \left(
	\ln \frac{1-\hat{C}/2+\sqrt{1-\hat{C}}}{1-\hat{C}+\sqrt{1-\hat{C}}}
	- \frac{\hat{C}}{2} \frac{\ln \frac{\hat{C}}{2-\hat{C}+2\sqrt{1-\hat{C}}}}{1-\hat{C}+\sqrt{1-\hat{C}}}
	\right) \nonumber \\
	&\simeq & - \mbox{Tr} \left( \frac{\hat{C}}{4} \ln \frac{\hat{C}}{4}\right)\ ,\label{eq:sent}
\end{eqnarray}
where the simpler form on the second line is valid when the eigenvalues of $\hat{C}$ are much smaller than one, as will be the case for us. 

Working to leading order in $K_{rs}$ as compared to the background $M_{rs}$, we can take the square root of $\kappa$ perturbatively in $K_{rs}$. Radial motion must lie in the $i,j,\ldots $ plane {\em or} the $a,b,\ldots$ plane because of the $SO(3)\times SO(6)$ symmetry of the background. $K_{rs}$, $M_{rs}$, and $C_{rs}$ are then $3\times 3$ or $6\times 6$; and $M$ is proportional to the identity matrix. We then have
\begin{eqnarray}
	\hat{\Omega} &&\simeq  \left(
	\begin{array}{cc}
		\sqrt{{M}} & 0 \\
		0 & \sqrt{{M}}
	\end{array}
	\right) \cdot \left[1 + 
		\left(
	\begin{array}{cc}
		0 & -\frac{K_{rs}}{M \sqrt{p_1^+p_2^+}} \\
		-\frac{K_{rs}}{M\sqrt{p_1^+p_2^+}} & 0
	\end{array}
	\right)\right]^{1/2} \nonumber \\
	&& \simeq \left(
	\begin{array}{cc}
		\sqrt{{M}} & -\frac{K_{rs}}{2\sqrt{M\,p_1^+p_2^+}} \\
		-\frac{K_{rs}}{2\sqrt{M\,p_1^+p_2^+}} & \sqrt{{M}}
	\end{array}
	\right)\label{eq:Omega}
\end{eqnarray}
where $M$ is either $\mu^2/9$ or $\mu^2/36$ given by~(\ref{eq:Mij}) or~(\ref{eq:Mab}). We have also dropped $\hat{K}$ in relation to $\hat{M}$ in the first matrix: to see this, first diagonalize $\hat{K}$, noting that $\hat{M}$ does not change since it is proportional to the identity; then drop eigenvalues of $\hat{K}$ in relation to $\hat{M}$ and retransform back. We then arrive at the $\hat{C}$ matrix
\begin{equation}\label{eq:Lambda}
	C_{rs} = \frac{1}{4\,M^2} \frac{K_{ru} K_{us}}{p_1^+p_2^+}
\end{equation}
where
\begin{equation}\label{eq:Ms}
	M = \frac{\mu^2}{9}\ \ \ \mbox{or}\ \ \ \frac{\mu^2}{36}\ .
\end{equation}
Along with~(\ref{eq:sent}), this gives us a measure of entanglement entropy between the two objects expressed in terms of the derivatives of the potential between them. $M$ comes from the PP-wave background and seem to function like a sort of `regulator'. 

Noting that, for radial motion, we must have 
\begin{equation}
	\dot{X}^r \propto X^r\ ,
\end{equation}
and looking at~(\ref{eq:Brs}) and~(\ref{eq:Krs}), we see that only fluctuations {\em parallel} to the relative velocity $\dot{X}^r$ contribute to the entanglement computation. Due to the spherical symmetry, the $\hat{K}$ and $\hat{C}$ matrices are effectively one by one. We take object 2 as the source and object 1 as the probe, in a regime where $P^+\equiv p^+_2\gg p^+_1 \equiv p^+$. And  using spherical coordinates where $\rho$ is the radial coordinate along the relative displacement vector between the two objects, {\em i.e.} $\rho^2 = X^r X^r$, we have $K_{\rho\rho}\equiv K$ with
\begin{equation}
	K = \partial_\rho\partial_\rho V -\partial_\rho\partial_\rho\delta_\rho V\, \dot{X}^\rho = -A^\rho\ .
\end{equation}
In the last step, we identified $K$ with the radial tidal acceleration as defined in~(\ref{eq:ALC1}) or~(\ref{eq:ALC2}).
We then have
\begin{equation}
	C = \frac{K^2}{4\,M^2\,p^+P^+}\ll 1
\end{equation}
since $K\ll M$. The Von Neumann entropy becomes
\begin{equation}\label{eq:finS}
	S_{ent} \simeq - \left( \frac{K^2}{16\,M^2\,p^+P^+} \ln \frac{K^2}{16\,M^2\,p^+P^+}\right)\ ,
\end{equation}
connecting quantum entanglement to the potential between the two bodies. In the next section, we will map $K$ onto supergravity geometry, completing the dictionary between entanglement and geometry through local tidal forces.

To conclude, we also use the potential given by~(\ref{eq:finpot}) to write an explicit equation for the entropy
\begin{eqnarray}
	K &=& \partial_{\rho}\partial_{\rho} V -\dot{\rho}\,\partial_{\rho}\partial_{\rho}\partial_{\dot{\rho}} V = \nonumber \\
	&=& p^+ P^+\begin{cases} \frac{13 \mu^4}{64\,R} \frac{1}{\rho^5} +\frac{ 5 \mu^2}{16\,R}\frac{1}{\rho^7} \dot{\rho}^2  + \frac{35 \mu^6}{20736\,R}\frac{(P^+)^2}{\rho^7} + \frac{305 \mu^6}{331 776\,R^3}\frac{1}{\rho^7} &\mbox{Motion in $SO(3)$} \\
	   -\frac{\mu^4}{64\,R}\frac{1}{\rho^5}+\frac{ 35 \mu^2}{16\,R}\frac{1}{\rho^7}\dot{\rho}^2  + \frac{35 \mu^6}{20736\,R}\frac{(P^+)^2}{\rho^7} + \frac{305 \mu^6}{331 776\,R^3}\frac{1}{\rho^7} &\mbox{Motion in $SO(6)$ } \end{cases}
	\label{eq:entex}\ .
\end{eqnarray}
Here $\rho^2 = (x^i)^2$ or $\rho^2=(x^a)^2$ depending on the case. The last two terms in each case arise from dipole membrane charge. Note that $M = \frac{\mu^2}{9}$ for $SO(3)$ and $M=\frac{\mu^2}{36}$ for $SO(6)$ so that both cases are divided by $\mu^2$ in~(\ref{eq:finS}). As an example, for the $SO(3)$ case, the entanglement between probe and source is, to leading order,
\begin{equation}
	\chi\equiv \frac{K}{4\,M\,\sqrt{p^+P^+}} \simeq -\frac{135 \mu^2}{256\,R} \frac{\sqrt{p^+ P^+}}{\rho^5} \Rightarrow S_{ent} \simeq -\chi^2 \ln \chi^2\ .
\end{equation}
The entanglement then decreases with larger distances between source and probe. Note that $\chi$ scales with the parameters of the BMN theory as $\mu^2/R^2$.

\vspace{0.5cm}{\bf Adiabatic regime}\vspace{0.5cm}

Before expanding the computation to the case with non-zero angular momentum, let us revisit a key assumption -- that the oscillator frequencies for the perturbations can be taken as time independent, {\em i.e.} we have quantum harmonic oscillators in an adiabatic regime. The adiabatic condition for a harmonic oscillator translates to
\begin{equation}\label{eq:adiab}
	\dot{\Omega} \ll \Omega^2
\end{equation}
where $\Omega$ is the frequency of the oscillator. Looking back at the case of radial motion and~(\ref{eq:Omega}), we see that
\begin{equation}
	\sqrt{\mbox{Tr}\,\dot{\Omega}^2} \sim \frac{\dot{K}}{\sqrt{M\,p_1^+p_2^+}}\ \ \ \mbox{and}\ \ \ \mbox{Tr}\,\Omega^2\sim {M}\ .
\end{equation}
This means that the condition~(\ref{eq:adiab}) can be satisfied provided we arrange
that 
\begin{equation}\label{eq:condad}
	\frac{\dot{K}}{M} \ll \sqrt{M\,p_1^+p_2^+}\ .
\end{equation}
Given the timescale of evolution in the PP-wave background would be set by the curvature scale $\sqrt{M}\sim \mu$, equation~(\ref{eq:condad}) becomes
\begin{equation}
    \frac{K}{M} \ll \sqrt{p_1^+p_2^+}
\end{equation}
which is satisfied when the probe is far enough away from the source. The adiabatic condition is then equivalent to the approximation scheme we have employed throughout -- that the source's back-reaction on the PP-wave background near the location of the probe is small. The case with non-zero angular momentum that we will treat next will not change this conclusion: we will see that adding the $B_{rs}$ couplings does not change things as the new relevant terms are damping terms. The general conclusion is that, given the way the $K_{rs}$ and $B_{rs}$ couplings arise in~(\ref{eq:Llong}), parametric resonance is avoided and the adiabatic regime holds as long as the probe is far enough from the source initially so that $K\ll M \sqrt{p_1^+p_2^+}$.

\vspace{0.5cm}{\bf Non-zero angular momentum}\vspace{0.5cm}

Consider next a probe moving around the source in a fixed two-dimensional plane which we denote as the $a$-$b$ plane, where $a$ and $b$ are fixed arbitrarily. Note that, given the $SO(3)\times SO(6)$ symmetry of the background, we would expect both $a$ and $b$ to lie in $SO(3)$ or $SO(6)$, but not straddle across. Furthermore, for the regime of interest, the distance between the source and the probe $X^r$ is to be large and hence the relative velocity would be small $\dot{X}^r$. In this scenario, we have all components of $B_{rs}=0$ {\em except}
\begin{equation}
	B_{ab}\neq 0\ \ \ \mbox{For motion in $a$-$b$ plane}
\end{equation}
Given that the classical trajectories lie in a plane $a$-$b$ and the general form of the potential is given by~(\ref{eq:finpot}), we see that both $K_{rs}$ and $B_{rs}$ are non-zero only in the $a$-$b$ subspace. The problem of finding the ground state of the perturbations in~(\ref{eq:Llong}) then becomes that of a charged particle in a magnetic field with an additional quadratic potential. The path integral is still gaussian and leads to a ground state wavefunction in $z$ that is that of simple harmonic oscillators -- except that the frequencies would be shifted by the magnetic field~\cite{Kleinert:2004ev}. All matrices $\hat{K}$, $\hat{B}$, and $\hat{C}$ are now four by four, two for the two particles times two from the planar motion, {\em i.e} the problem lives in the space of $z_{1,2}^{a,b}$. We divide this space into a direct product of the $1$-$2$ particle space and $a$-$b$ Lorentz space. Going to Fourier space, with $\omega$ representing the Fourier frequency dual to the light-cone time and using a tilde as in $\tilde{z}$ to denote Fourier modes, we write the Lagrangian as
\begin{equation}
	L= -\frac{\omega^2}{2} {\tilde{z}}^*\cdot {\tilde{z}}-\frac{1}{2} \tilde{z}^*\cdot \hat{\kappa} \cdot \tilde{z}  - \frac{i\ \omega}{2} \tilde{z}^* \cdot \hat{\mathcal{B}} \cdot {\tilde{z}}
\end{equation}
where $\tilde{z}$ are the Fourier modes. Diagonalize $K_{ab}$ so that 
\begin{equation}
	K_{ab} \rightarrow  \left(
	\begin{array}{cc}
		k & 0 \\ 0 & k'
	\end{array}
	\right)
\end{equation}
and
\begin{equation}
	B_{ab} \rightarrow  \left(
	\begin{array}{cc}
		0 & B' \\ -B' & 0
	\end{array}
	\right)
\end{equation}
while $M$ is unchanged since it is proportional to the identity matrix in the $a$-$b$ space (in $SO(3)$, we have $M=\mu^2/9$, and in $SO(6)$, $M=\mu^2/36$). We also have absorbed a factor of ${1}/{\sqrt{p_1^+p_2^+}}$ in $B'$, $k$, $k'$ to keep the notation clean. We then write
\begin{equation}
	\hat{\mathcal{B}} = \left(
	\begin{array}{cc}
		0 & B'_{ab} \\ -B'_{ab} & 0
	\end{array}
	\right) \otimes \left(
	\begin{array}{cc}
		0 & -1 \\ -1 & 0
	\end{array}
	\right)  
\end{equation}
and 
\begin{equation}
	\hat{\kappa} = M\otimes\left(
	\begin{array}{cc}
		{1} & 0 \\
		0 & {1}
	\end{array}
	\right) + 
		\left(
	\begin{array}{cc}
		k & 0 \\ 0 & k'
	\end{array}
	\right)\otimes\left(
	\begin{array}{cc}
		0 & -1 \\
		-1 & 0
	\end{array}
	\right)
\end{equation}
The general solution involves diagonalization of a four by four matrix
\begin{equation}
	L = -\frac{1}{2} \tilde{z}^* \cdot\left(
	\begin{array}{cccc}
		\omega^2+{M} & -k & 0 & -i\omega B' \\
		 - k & \omega^2+{M} & -i\omega B' & 0 \\
		0 & i\omega B' & \omega^2+{M} &  - k' \\
		i\omega B' & 0 &  -k' & \omega^2+{M}
	\end{array}
	\right)\cdot \tilde{z}\ .
\end{equation}
For a giant graviton or metastable star back-reacting on the PP-wave geometry, the $a$-$b$ subspace is necessarily isotropic and hence $k=k'$; the matrix is then easily diagonalized in terms of  
\begin{eqnarray}
	&&\tilde{y}_1=(\tilde{z}^b_1+\tilde{z}^b_2)+i (\tilde{z}^a_1+\tilde{z}^a_2)\ \ \ ,\ \ \ \tilde{y}_2=-(\tilde{z}^b_1-\tilde{z}^b_2)+i (\tilde{z}^a_1-\tilde{z}^a_2) \nonumber \\
	&&\tilde{y}_3=(\tilde{z}^b_1+\tilde{z}^b_2)-i (\tilde{z}^a_1+ \tilde{z}^a_2)\ \ \ ,\ \ \ \tilde{y}_4=- (\tilde{z}^b_1-\tilde{z}^b_2)-i (\tilde{z}^a_1- \tilde{z}^a_2)\label{eq:evectors}
\end{eqnarray}
with eigenvalues $\omega^2 \pm \omega\, B' + M  \pm' k$, or written more suggestively
\begin{eqnarray}
	&&\left(\omega -\frac{1}{2}\left(\pm B' - \sqrt{{B'}^2\pm'4\,k -4\,M}\right)\right)\left(\omega -\frac{1}{2}\left(\pm B' + \sqrt{{B'}^2\pm'4\,k -4\,M}\right)\right) \nonumber \\
	&&\simeq \left(\omega \pm \frac{B'}{2} \pm' \frac{i}{2}\frac{k}{\sqrt{M}} - i \sqrt{M}\right)\left(\omega \pm \frac{B'}{2} \mp' \frac{i}{2}\frac{k}{\sqrt{M}} + i \sqrt{M}\right)\ ,
\end{eqnarray}
where in the second line we expanded for $k, B' \ll M$ to leading order.
The important point is that the eigenvectors are independent of $B'$. This leads to a diagonalized action of the form
\begin{equation}\label{eq:shifted}
	\frac{1}{2}\left[\left(- i \partial_+ - \frac{B'}{2} + \frac{i}{2}\frac{k}{\sqrt{M}} - i \sqrt{M}\right)y_1\right] \left[\left(i \partial_+ - \frac{B'}{2} - \frac{i}{2}\frac{k}{\sqrt{M}} + i \sqrt{M}\right) y_1\right] + \cdots
\end{equation}
with the dots denoting three additional terms for the eigenvectors $y_2$, $y_3$, and $y_4$ corresponding to flipping the signs of $B'$ and $k$ independently. This implies that, in constructing the $\hat{\Omega}$ matrix in~(\ref{eq:omega}) -- after transforming back to the $z$ basis using~(\ref{eq:evectors}), $B'$ will appear only at quadratic order -- which is sub-leading and must be dropped for consistency (the cross terms with $B'$ in~(\ref{eq:shifted}) cancel). 
The $\hat{C}$ matrix in~(\ref{eq:lambda}) from which the entanglement entropy is constructed will then not have dependence on $B'$. The relation we found for the radial scenario earlier is hence unchanged. The reason this is interesting is twofold: (1) The entropy-potential relation given by~(\ref{eq:sent}) and~(\ref{eq:Lambda}) then holds irrespective whether the two entangled objects are orbiting each other or falling radially; (2) We have not been able to map $B_{rs}$ from~(\ref{eq:Brs}) onto an obvious geometrical quantity in supergravity, unlike $K_{rs}$ as we shall soon see. If $B_{rs}$ was to appear in the entanglement entropy, it would seem to imply a `non-geometrical' component to the entanglement entropy -- which apparently does not happen.

\section{A covariant form}

In this section, we want to try to rewrite the relation established between entropy and potential by~(\ref{eq:sent}), (\ref{eq:Lambda}), and (\ref{eq:Ms}) into a covariant form that relates entropy and geometry directly. The key link we used to map entropy onto interaction potential involved the tidal acceleration. In general, if a probe is moving in some background geometry, the tidal acceleration it experiences is given by~\cite{waldbook,Kouretsis:2010nu,Ellis:1984bqf}
\begin{equation}\label{eq:Agen}
	A^\mu = u^\nu \nabla_\nu \left(u^\rho \nabla_\rho \xi^\mu\right) = -R_{\rho\nu\lambda}^{\ \ \ \ \mu} u^\rho u^\lambda \xi^\nu + \xi^\nu \nabla_\nu\left(a^\mu\right)
\end{equation}
where $u^\mu$ is the probe's velocity, and $\xi$ is a space-like vector transverse to this velocity $\xi^\mu u_\mu=0$. $a^\mu$ is the probe's acceleration and would be zero if it was subject to only gravitational forces -- $a^\mu=0$ is simply the geodesic equation; but $a^\mu \neq 0$ for example due to additional forces arising from membrane dipole charge. Note that the tidal acceleration is related to the second characteristic form of the geometry and hence to the rate at which areas transverse to the motion are shrinking, expanding, and twisting\footnote{For a congruence of geodesics with velocity $u^\mu$, the second characteristic form is given by
\begin{equation}
	B_{\mu\nu} = \nabla_\mu u_\nu\ .
\end{equation}
$B_{\mu\nu}$ describes how a transverse area to the velocity squeezes, rotates, and twists as you move along the geodesic. Correspondingly, tidal acceleration is 
\begin{equation}
	a^\mu = u^\nu \nabla_\nu \left(B^\mu_{\ \rho}\xi^\rho\right)
\end{equation}
where $\xi^\mu u_\mu = 0$. 
}.
In our previous discussion, we encountered the tidal acceleration given by 
\begin{equation}
	A^r = -K_{sr} \xi^s\ .
\end{equation}
Note that $A^r$ includes subtracting the effect of the background PP-wave geometry. The non-zero components of the Riemann tensor of the PP-wave background are given by $R^{PP}_{+r+s} = M_{rs}$. Using a minimal prescription to generalize these equations to covariant form, we have
\begin{equation}\label{eq:map}
	{K_{rs}}=(R_{\mu s \nu r}-R_{\mu s \nu r}^{PP}) u^\mu u^\nu - \nabla_s a_r\ .
\end{equation}
Note the dependence on the probe's velocity: this makes intuitive sense as this is entanglement seen from the perspective of the probe.
We then write
\begin{equation}
	C_{rs} = \frac{1}{4\,\beta^2} {K_{ru} K_{us}}
\end{equation}
with $S(C)$ given by~(\ref{eq:sent})
\begin{equation}
	S_{ent} \simeq - \mbox{Tr} \left( \frac{\hat{C}}{4} \ln \frac{\hat{C}}{4}\right)\ .
\end{equation}
and defining 
\begin{equation}
	\beta^2 = (R^{PP}_{\mu v\nu w}u^\mu u^\nu)^2
\end{equation}
One key observation of this relation is that this entanglement entropy, constructed from $C$, is necessarily aware of the effects of {\em all} supergravity fields, metric {\em and} 4-form flux, as can be seen from the second term on the right side of equation~(\ref{eq:map}). Note also that in the absence of the source, $K_{rs}\rightarrow 0 \Rightarrow S_{ent}\rightarrow 0$ as needed. We also note that this entropy is very small -- which is a reflection of the fact that the local curvature scale at the probe is to be much less than that of the background PP-wave curvature scale. In the flat space case treated in~\cite{Sahakian:2019cxc}, the dimensionless $\hat{C}$ relates to the ratio of the local curvature scale to the Planck scale. So, generically this entropy is to be very small as a statement that gravity is a weak force. It is however particularly interesting to explore emergence of spacetime from entropy in regimes where gravity is strong. This, however, is beyond the regime of validity of our computations.

\section{Conclusion}\label{sec:conclusion}

Entanglement entropy for a physical system can be defined in a myriad of ways -- and most importantly depends on how one slices the physical system. We have shown that there is a certain measure of entanglement between two interacting objects that lends itself to a geometrical interpretation. The proposal is particularly interesting and unique because it connects a physical quantity that is inherently non-local, entanglement, to local spacetime geometry. If it is general, it underscores the emergent geometry paradigm -- that spacetime geometry is an approximate emergent phenomenon arising from quantum entanglement.

The question is then how general is our treatment. We tried to present the narrative starting with broad strokes, but then gradually narrowing onto the specific scenario of two giant gravitons or metastable stars in eleven dimensional supergravity in a PP-wave background; one object was to be much larger than the other, and the smaller one was far enough away to function as a probe. This more specific setup was necessary to make the computation tractable. However, there were also a series of assumptions that were critical to the map between entanglement and geometry, to delineating the regime in which quantum entanglement acquires a geometrical character:

\begin{itemize}
	\item We needed that the distance between source and probe be large enough so that the source back-reacts weakly on the PP-wave background at the location of the probe. This allowed us to employ a perturbative expansion, and, at first, seems to be for computational convenience only. However, in BMN language, this regime allows us to integrate out off-diagonal matrix modes, correlating with the off-diagonal modes being heavy. This is a key benchmark for the emergence of geometry from matrix quantum degrees of freedom. If all matrix degrees of freedom in BMN theory are to participate in the dynamics, we expect no dual geometrical picture; only in the regime where off-diagonal modes get frozen can we expect the emergence of the geometrical picture -- the imprint of which in the matrices lies in the dynamics of the diagonal modes. 
	\item We needed to employ an adiabatic regime, where the classical evolution of the two interacting objects happens on a time scale that is much longer than that of small fluctuations from the classical paths. We demonstrated however that this regime is equivalent to the previous one: large enough distances between source and probe.
	\item We used a source that conforms to $SO(3)\times SO(6)$ symmetry of the PP-wave background -- a giant graviton or a metastable star. This is needed to control the computation, to make the equations more symmetric. It is not clear whether this assumption plays a fundamental role, or if it is just a technical convenience as is typical in physics when one harnesses the benefits of symmetry.
	\item The setup was implemented in a PP-wave background, corresponding to BMN theory on the dual side. This allowed us to use stable or metastable matrix configurations for the interacting bodies and avoid potential red herrings. This also appears to have served as a sort of `regulator' in the final entropy expression. However, a similar map between entanglement and geometry was already worked out in~\cite{Kabat:1997im} for Matrix theory in flat-space, without the PP-wave background. This suggests that the PP-wave setting was mostly a computational convenience.    
	\item Working in the Light-cone frame seems to have played an important role, as we shall argue below. The setting allows us to define a natural $D-2$-dimensional local space-like subspace transverse to the probe's velocity, where $D$ is spacetime dimension. In a sense, this leads to a kind of holography where the role of a `holographic screen' is played by this transverse space -- transverse to the light-cone direction {\em and} the velocity of a local probe. 
\end{itemize}

Before we try to generalize the entropy-geometry relation, let us emphasize a few of the non-trivial aspects of the specific relation we arrived at:
\begin{itemize}
	\item The connection between geometry and entanglement through tidal acceleration was presented from the gravity side of the duality. At first, it might seem strange that the entropy-geometry relation can be made entirely in the supergravity context. But we must remember that the computation of the entanglement could be performed entirely in a non-gravitational framework, in BMN Matrix theory, with the same results, since the gravitational framework is dual to BMN theory. Indeed, in~\cite{Sahakian:2019cxc}, this was done in the case of the BFSS Matrix theory. Presenting the computation from the dual gravity perspective underscores how generic is the phenomenon: {\em as long as a Matrix theory computation lies in a realm where a gravitational dual description is valid, the map between entanglement and geometry holds}. The map can break down only in a regime of Matrix theory that does not have a well-defined dual gravitational description. In such a regime, the entanglement entropy would still be an observable, but it would not have a geometrical character.
	\item The relation seems to be independent of the relative motion between probe and source. And a problematic non-geometrical piece that we called $B_{rs}$ dropped out from the final result. This is a testament to the robustness of the treatment.
	\item The entanglement entropy we computed included contributions from non-gravitational interactions in supergravity -- membrane dipole interactions. Generally, we would expect that the full supergravity field content can participate in the `geometrization' of the entanglement entropy.
\end{itemize}

How general can the relation for the cases of two interacting spherical objects in a PP-wave background in eleven dimensional supergravity be? To answer this question, we will need to engage in a certain level of speculation -- albeit guided by symmetries and the results we have. Let us imagine that we do not have the PP-wave background, but we start in flat space light-cone supergravity. Then the key ingredient from which the entanglement between two interacting objects can be built from the tensor
\begin{equation}
	r_{rs} = \frac{1}{2\,\beta}(R_{\mu s \nu r}-R_{\mu s \nu r}^{PP}) u^\mu u^\nu \rightarrow R_{\mu s \nu r} p^\mu p^\nu\ .
\end{equation}
This is dimensionless, replacing the scale $\mu$ (hidden in $\beta$) for the case of a PP-wave background with $p^\mu$ which represents the momentum of the probe -- the only relevant scale when in flat space. For simplicity, we also dropped interactions due to membrane charge. This $r_{rs}$ tensor is the central building block and lives in $D-2=9$ dimensional space transverse to the momentum of the probe. In cosmology, there is a convenient $1+3$ covariant description of general relativity where one defines\cite{Tsagas:2007yx}
\begin{equation}
	h_{\mu\nu} = g_{\mu\nu} + u_\mu u_\nu
\end{equation}
which allows a unique decomposition of every spacetime quantity into its irreducible time-like and space-like components -- along the velocity and transverse to it. We want to do something similar, however specialize to the light-cone framework we are working with. In the light-cone frame of supergravity we are considering, all objects carry fixed longitudinal momenta along the light-cone direction. Transverse means transverse to the light-cone direction {\em and} the velocity of the probe. We then define the projector
\begin{equation}
	\xi^\mu_r = \delta^\mu_r - \frac{p_r}{p_-}\delta^\mu_-\ \Rightarrow \xi^+_r = 0
\end{equation}
where $r=1,\ldots , D-2$. We can then write
\begin{equation}
	r_{rs} = \xi^\alpha_r \xi^\beta_s R_{\mu \beta \nu \alpha} p^\mu p^\nu\ .
\end{equation}
This object references only the local geometry seen by the probe and its momentum. From this, we are to construct tensor
\begin{equation}\label{eq:ctensor}
	C_{rs} = {r_{ru} r_{us}}	
\end{equation}
which we will call suggestively the {\em `c-tensor'}. Then we conjecture that there is a certain entanglement entropy between probe and environment that is related to local geometry experienced by the probe, and it is given by 
\begin{equation}\label{eq:finalS}
	S_{ent} = - \mbox{Tr} \left( \frac{\hat{C}}{4} \ln \frac{\hat{C}}{4}\right)\ .
\end{equation}
Beside being closely guided by the example of gravitating giant gravitons in BMN theory, there is something peculiar about this relation: the trace of $r_{rs}$ is related to the energy-momentum measured by the probe locally, and this energy-momentum is responsible for the local curvature through Einstein's equations. Equation~(\ref{eq:ctensor}) looks reminiscent of an OPE of two energy-momenta tensors relating to a central charge in a conformal theory -- with $\xi$ playing the role of distance between nearby geodesics. And the entropy relation~(\ref{eq:finalS}) is itself suggestive of entanglement entropy with $\hat{C}$ being a sort of central charge. Of course, there is no obvious conformal field theory here, and these similarities are just heuristic -- but they do suggest that the entanglement entropy relation we derived in BMN theory might be part of a more general holographic relation.

Our conjecture needs to be tested further in two main directions. First, can one find other computationally accessible settings in Matrix theories where this entanglement entropy can be computed, with the suggested relation to geometry verified or falsified? Second, if the general relation holds, it might be possible to extract gravitational dynamics from the quantum mechanics of interacting matrix degrees of freedom; that is, can one use knowledge of the general properties of entanglement entropy to arrive at constraints on the Riemann tensor of the emergent geometry that amount to the Einstein equations, as in for example~\cite{Jacobson:1995ab}?

\newpage
\section{Acknowledgments}

VS would like to thank the IPhT at Saclay for hosting him. This work was supported by NSF grant number PHY-0968726. We thank Adam Busis for participating in the initial stages of the BMN computation.

\bibliographystyle{utphys}
\providecommand{\href}[2]{#2}\begingroup\raggedright\endgroup

\end{document}